# Detection of VHE γ-rays from Mkn 421 with the HEGRA Cherenkov Telescopes

D. Petry[1], S.M. Bradbury[1], A. Konopelko[2], J. Fernandez[1,3], F. Aharonian[2], A.G. Akhperjanian[4], A.S. Belgarian[4], J.J.G. Beteta[3], J.L. Contreras[3], J. Cortina[3], A. Daum[2], T. Deckers[5], E. Feigl[1], V. Fonseca[3], B. Funk[6], J.C. Gonzalez[3], V. Haustein[7], G. Heinzelmann[7], G. Hermann[2], M. Hess[2], A. Heusler[2], W. Hofmann[2], R. Kankanian[2,4], O. Kirstein[5], C. Köhler[2], H. Krawczynski[7], H. Kornmayer[1], A. Lindner[7], E. Lorenz[1], N. Magnussen[6], H. Meyer[6], R. Mirzoyan[1,3,4], H. Möller[6], A. Moralejo[3], N. Müller[5], L. Padilla[3], M. Panter[2], R. Plaga[1], J. Prahl[7], C. Prosch[1], G. Rauterberg[5], W. Rhode[6], V. Sahakian[4], M. Samorski[5], J.A. Sanchez[3], D. Schmele[7], W. Stamm[5], M. Ulrich[2], H.J. Völk[2], S. Westerhoff[6], B. Wiebel-Sooth[6], C.A. Wiedner[2], M. Willmer[5], and H. Wirth[2]

[1] Max-Planck-Institut für Physik, Föhringer Ring 6, D-80805 München, Germany
[2] Max-Planck-Institut für Kernphysik, P.O. Box 103980, D-69029 Heidelberg, Germany
[3] Facultad de Ciencias Fisicas, Universidad Complutense, E-28040 Madrid, Spain
[4] Yerevan Physics Institute, Yerevan, Armenia
[5] Universität Kiel, Inst. für Kernphysik, Olshausenstr.40, D-24118 Kiel, Germany
[6] BUGH Wuppertal, Fachbereich Physik, Gaußstr.20, D-42119 Wuppertal, Germany
[7] Universität Hamburg, II. Inst. für Experimentalphysik, Luruper Chaussee 149, D-22761 Hamburg, Germany

**Abstract.** A detection of γ-rays from Mkn 421 at energies above 1 TeV is reported, based on observations made in December 1994 - May 1995 with the first two HEGRA Cherenkov telescopes. From the image analysis, 111 excess gamma candidates are obtained from the 26 h Telescope #1 (CT1) dataset (significance 4.0 $\sigma$) and 218 from the 41 h Telescope #2 (CT2) dataset (significance 4.2 $\sigma$) at zenith angles $\theta < 25°$. The combined significance is approx. 5.8 $\sigma$. This is the second detection of Mkn 421 at TeV energies. The average excess rate is $4.3 \pm 1.0$ h$^{-1}$ for CT1 and $5.4 \pm 1.3$ h$^{-1}$ for CT2. Comparison with our contemporary observations of the Crab Nebula indicates that Mkn 421 has a steeper spectrum than the Crab Nebula above 1 TeV. Under the assumption that the spectrum of Mkn 421 follows a power law, we obtain a differential spectral index of $3.6 \pm 1.0$ and an integral flux above 1 TeV of $8(\pm 2)_{\mathrm{Stat}}(+6-3)_{\mathrm{Syst}} \times 10^{-12}$ cm$^{-2}$s$^{-1}$ from a comparison with Monte Carlo data. This flux is smaller than the Crab Nebula flux by a factor of $2.0 \pm 0.8$.

## 1. Introduction

Being the nearest BL Lac object (redshift $z = 0.03$), Mkn 421 has been studied by many and in all wavebands.

EGRET, onboard the Compton Gamma Ray Observatory (CGRO), detected it in the 50 MeV to 10 GeV range (Lin et al. 1992). Shortly after, the Whipple group observed γ radiation from Mkn 421 between 0.5 and 1.5 TeV at a confidence level of 6.3 $\sigma$ (Punch et al. 1992). The Whipple group has since then routinely monitored the Mkn 421 flux which appears highly variable on timescales of several months to days (Kerrick et al. 1995, Macomb et al. 1995). For the period of time covered by this work, Mkn 421 was in a high state according to the Whipple group (Lamb et al. 1995).

The HEGRA collaboration's imaging air Cherenkov telescopes are part of its cosmic ray detector complex (e.g. Fonseca et al. 1995) at the Observatorio del Roque de los Muchachos on the Canary Island of La Palma (28.75° N, 17.89° W, 2200 m a.s.l.). The first two telescopes CT1 and CT2 are described elsewhere (Mirzoyan et al. 1994; Wiedner 1994 and Rauterberg et al. 1995). They are 93 m apart and have energy thresholds of approx. 1.2 TeV and 1.0 TeV respectively.

## 2. Observations and data analysis

Here, alongside our results, we merely outline the data reduction procedure; for further details and results for zenith angles $\theta > 25°$ see Petry et al. (1995) and Petry (1996). Note that in this letter the determination of the effective collection area has been improved since Petry et al. (1995).

*2.1. Selection and analysis of data from Telescope 2*

Between December 1994 and May 1995 we recorded 91 hours of ON-source observations of Mkn 421 with CT2 at zenith angles between 9° and 55°. In a 23 minutes ON/OFF or OFF/ON rhythm, matching OFF-source observations were made. From these, 161 ON/OFF pairs were selected on the basis of weather information recorded by the observers and extinction measurements from the Carlsberg Automatic Meridian Circle near the HEGRA site. They correspond to 679098 events ON and 678293 events OFF. These data were analysed in three steps: (1) flat-fielding and calibration, (2) filtering to obtain a dataset of showers with well determined image parameters (for a definition see e.g. Reynolds et al. 1993), (3) selection of gamma candidates using cuts on the image parameters.

1. Calibration and flat fielding are based on regular measurements of the pedestals and the relative photomultiplier gains.
2. In addition to a hardware trigger condition of 2 out of 61 pixels fired (including at least 1 of the central 37) a software trigger condition of 2 out of 37 were used to exclude camera-edge events with incomplete images. A CONC<0.95 cut and data from an anticoincidence shield was used to reject events due to cosmic ray muons. Events recorded under poor telescope positioning were rejected leaving a mean absolute pointing error of 0.1° for events after filter. Each observation was scanned for short-term rates significantly above or below the mean rate of that run and run pairs containing abnormal peaks were removed.
3. In order to enhance the signal, a series of image parameter cuts was applied which enrich the sample with γ-shower candidates. These cuts were optimized on Monte Carlo simulations and real data (30/30 hours of ON/OFF observations of the Crab Nebula made with CT2 in 1994 at $\theta < 30°$).

$$0.6° < \text{DIST} < 1.2°$$
$$0.06° < \text{WIDTH} < 0.2°$$
$$0.18° < \text{LENGTH} < 0.375°$$
$$0.5 < \text{CONC}$$
$$\text{ALPHA} < 15°$$

*2.2. Selection and analysis of data from Telescope 1*

In December 1994 CT1 was equipped with a new high resolution 127 pixel camera (Rauterberg et al. 1995) with high QE photomultipliers and hollow light cones. 101 ON/OFF pairs (38.9 h ON/ 38.9 h OFF) recorded between January and May 1995 were accepted for analysis (189779 events ON and 191496 events OFF).

The flat fielding and filtering of the data was done in a way similar to that for CT2. A software trigger 2/91 was applied to the calibrated signals. The mean absolute pointing error after filter was 0.1°.

Since for the new camera neither observations of the Crab nebula nor all parameters for the Monte Carlo optimization of the cuts were available, we used the set of cuts developed for a 91 pixel camera with similar resolution as described in Reynolds et al. (1993):

$$0.51° < \text{DIST} < 1.1°$$
$$0.07° < \text{WIDTH} < 0.15°$$
$$0.16° < \text{LENGTH} < 0.30°$$
$$\text{ALPHA} < 15°$$

In addition, a cut 0.4 < CONC was applied analogous to the CONC cut for CT2 and taking into account the higher resolution of the camera. The investigation of the efficiency of these cuts when applied to CT1 data is underway. In this letter we present flux calculations for CT2 only.

## 3. Results

Table 1 shows the event statistics for both telescopes for a $\theta < 25°$; from the CT2 data we determine spectral index and flux. In order to compensate for possible variations in the night sky background, the excesses $N_{ex}$ after cuts were calculated using $N_{ex} = N_{ON} - \frac{\nu_{ON}}{\nu_{OFF}} N_{OFF}$ where $\nu$ is the number of events with $35° < \text{ALPHA} < 85°$ in which region no source gamma candidates are expected. The statistical significances are calculated according to Li & Ma (1983) Eq. 9.

For comparison, 49.5 h of observations of the Crab Nebula taken with CT2 between September 1994 and March 1995 at $\theta < 25°$ were analysed using the same procedure as described here for Mkn 421. In agreement with our earlier publication (Konopelko et al. 1996) we obtain with a zenith angle cut at 25° an excess of 709 events with a significance of 9.3 $\sigma$.

The ALPHA distributions for Mkn 421 and the Crab Nebula are shown in figure 1.

## 4. Spectral Index and Flux calculations for CT2 data

By comparison with MC data produced using code described in Konopelko et al. (1996) it was possible to estimate the spectral index and the integral flux for Mkn 421 above 1 TeV from the CT2 data.

The MC databank used here consists of 28k (170k) simulated γ (proton) showers for $\theta = 0°$ and 19k (45k) γ (proton) showers for $\theta = 30°$ with primary energies between 0.05 and 30 TeV. The telescope simulation comprised the generation of Poisson Night Sky Background noise (NSB) similar to the measured background and a simple simulation of the mirror imperfections performed by convoluting the camera image with a two-dimensional Gaussian with a FWHM equal to the measured point spread function of the telescope of 0.10° (value at 1° off axis).

The effective collection area of CT2 for the given set of cuts was calculated taking into account the dependence

| Results for Mkn 421 from CT1 and CT2, zenith angle $< 25°$ | | | | | | | | | | |
|---|---|---|---|---|---|---|---|---|---|---|
| | Time (h) | | | | events remaining | | | | excess (see text) | | significance | |
| | CT1 | | CT2 | | CT1 | | CT2 | | | | | |
| analysis stage | ON | OFF | ON | OFF | ON | OFF | ON | OFF | CT1 | CT2 | CT1 | CT2 |
| after filter | 25.62 | 25.65 | 40.51 | 40.59 | 81270 | 80641 | 240571 | 240694 | 723 | 351 | 1.8 σ | 0.5 σ |
| after all cuts except ALPHA | " | " | " | " | 1904 | 1757 | 6945 | 6635 | 149 | 204 | 2.5 σ | 1.7 σ |
| **after all cuts** | " | " | " | " | **451** | **340** | **1448** | **1213** | **111** | **218** | **4.0 σ** | **4.2 σ** |

**Table 1.** Event statistics for the cuts described in section 2 for the data from both telescopes.

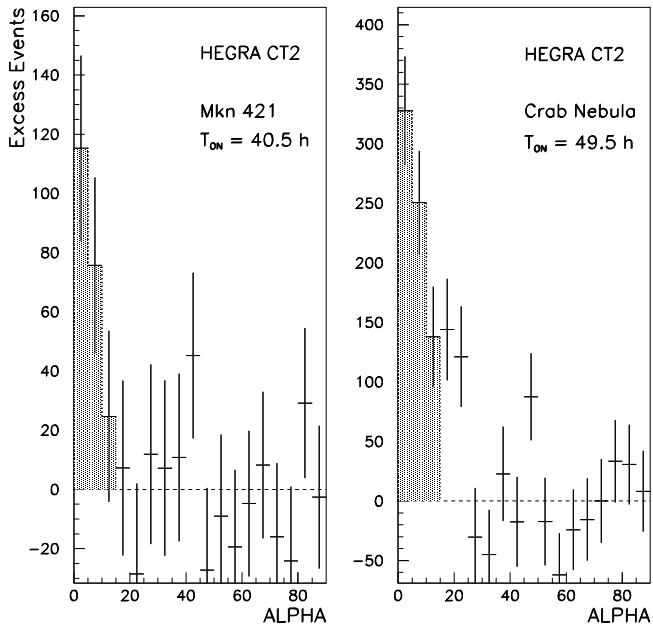

**Fig. 1.** The ON-source - OFF-source ALPHA distributions for CT2 data with $\theta < 25°$. For Mkn 421, there are 218 excess gamma candidates (shaded, rate 5.4 h$^{-1}$) with a significance of 4.2 σ, for the Crab Nebula there are 709 excess gamma candidates (rate 14.3 h$^{-1}$) with a significance of 9.3 σ.

of the cut efficiencies on primary energy and impact parameter. The effective collection area is strongly energy dependent. It is therefore impossible to determine an integral flux without prior determination of the shape of the source spectrum. With a given source spectrum, the flux can be determined by convoluting the spectrum with the effective collection area as a function of primary energy giving an expected integral rate for comparison with the observed rate (see Aharonian et al. 1995). The flux values given below have two main uncertainties:

Firstly, the absolute energy calibration of CT2: In Konopelko et al. (1996) we obtained a conversion factor $\lambda$ = 1.5 ± 0.4 ADC counts per photoelectron from adjusting the all particle cosmic ray rate expected from the Monte Carlo simulation to the measured rate. However, using a method presented in Mirzoyan et al. (1995), the conversion factor was measured from single photoelectron spectra to be 1.05 ± 0.1 ADC counts per photoelectron.

Until additional measurements resolve these problems, we now consider $\lambda$ to have the nominal value 1.25±0.25 ADC counts per photoelectron.

Secondly, due to the low significance of the Mkn 421 signal, the rather coarse energy resolution of the telescope and the integration over many experimental parameters (zenith angle, distance between shower axis and optical axis of the telescope, variations of the atmospheric transmission), the shape of the spectrum cannot be determined with high accuracy. Under the simple assumption that the spectrum follows a power law, two different methods were used to examine the spectra of both Mkn 421 and the Crab Nebula:

1. A power law fit to the number of excess events after all cuts as a function of the energy threshold $E_{thr}$. We obtain directly from the fit the *integral* spectral indices 2.8±0.6 for Mkn 421 and 1.8±0.2 for the Crab Nebula. These are only very approximate as the energy dependence of the image cut efficiencies is ignored. Still this is an indication that the Mkn 421 spectrum is steeper than that of the Crab Nebula above 1 TeV.

2. Comparison of the observed and simulated distributions of the total amount of light in the image (image parameter SIZE). This procedure is illustrated in figure 2. The distribution for γ events was simulated for differential spectral indices of 1.5, 1.75, ..., 4.5. A conversion factor $\lambda$ = 1.25 ADC counts per photoelectron was assumed and the NSB and the mean zenith angle of the dataset were taken into account. A logarithmic binning with 4 bins between the average SIZE corresponding to 1.0 TeV and to 10 TeV was used (figure 2b). These distributions were then compared to the experimental SIZE distribution for excess γ-candidates obtained by subtracting ON - OFF (figure 2a). The resulting reduced $\chi^2$ values as a function of the spectral index are shown in figure 2c. The function is well approximated by a parabola near the minimum from which we calculate the most probable spectral index under the assumption that the source has a power law spectrum (in accordance with the minimum reduced $\chi^2 < 1$). We obtain *differential* spectral indices of 3.6±1.0 for Mkn 421 and 2.7±0.3 for the Crab Nebula. Note the large error on the spectral index of Mkn 421 due to the low significance of the signal. The proce-dure was repeated for the conversion factors $\lambda = 1.0$ (1.5) ADC counts per photoelectron which resulted in indices shifted by -0.1 (+0.1) for both sources.

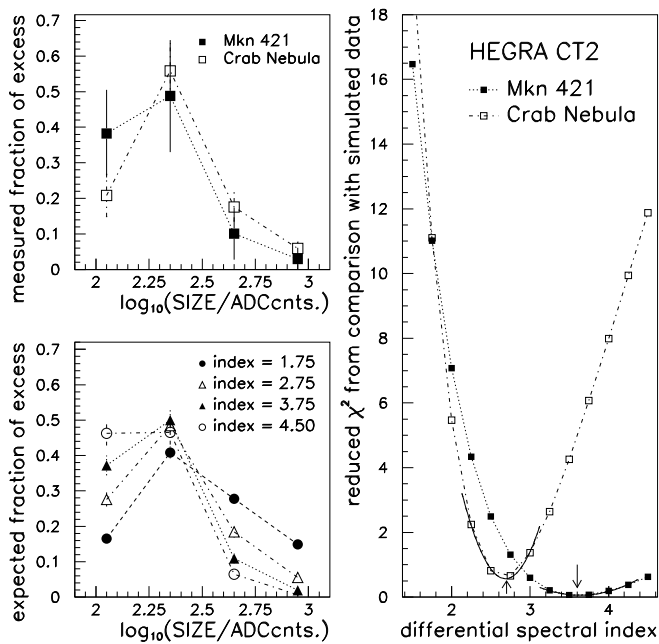

**Fig. 2. a** (upper left) the measured differential distributions of the image parameter SIZE (difference ON - OFF). SIZE is the total amount of light in the shower image (here in ADC counts for CT2: 1.25±0.25 ADC counts = 1 photoelectron). This is a instrument dependent measure for the primary energy (e.g. for zenith angle = 0° and E = 1 TeV, $\log_{10}(\text{SIZE}) \approx 2$ on average; for E = 10 TeV, $\log_{10}(\text{SIZE}) \approx 3$ on average). **b** (lower left) examples of simulated SIZE distributions for four different differential spectral indices of the γ-ray source. **c** (right) the reduced $\chi^2$ values from the comparison of the measured distributions for Mkn 421 and the Crab Nebula with simulated SIZE distributions for 13 different spectral indices (see text).

Based on the CT2 observations at $\theta < 25°$ which gave an excess rate of 5.4 h$^{-1}$, the mean flux above 1 TeV from Mkn 421 in the period from 29 December 1994 to 5 May 1995 was calculated assuming a differential spectral index of 3.6±1.0 to be

$$0.8(\pm 0.2)_{\text{Stat}}(+0.6 - 0.3)_{\text{Syst}} \times 10^{-11} \text{ cm}^{-2}\text{s}^{-1}$$

The figures in brackets are the statistical error of the rate and the systematic errors due to the stated uncertainties in the energy calibration and the range of spectral indices. For the signal from the Crab Nebula (rate = 14.3 h$^{-1}$) we find the flux above 1 TeV with a differential spectral index of 2.7±0.3 to be

$$1.5(\pm 0.2)_{\text{Stat}}(+1.0 - 0.5)_{\text{Syst}} \times 10^{-11} \text{ cm}^{-2}\text{s}^{-1}$$

This Crab Nebula flux is larger than the value given in our earlier publication (Konopelko et al. 1996) because this analysis uses a smaller factor for the conversion from photoelectrons to ADC counts and includes imperfections of the telescope optics in the Monte Carlo simulation.

## 5. Conclusion

We report a detection of γ-radiation from Mkn 421 at energies above 1 TeV using two imaging air Cherenkov telescopes. An integral flux above 1 TeV of approx. $0.8 \times 10^{-11}$ cm$^{-2}$s$^{-1}$ has been measured from the December 1994 - May 1995 CT2 data. Mkn 421 shows a spectrum which is compatible with a differential spectral index of 3.6±1.0. This result is not inconsistent with the Whipple group's determination of the energy spectrum of Mkn 421 (Mohanty et al. 1993) (which is, however, not contemporary) and is in fact additional evidence of a spectral steepening for Mkn 421 above 1 TeV (e.g. Biller et al. 1995 and references therein). With the completion of the HEGRA 5-telescope array in winter 1996/97, a detailed investigation of the spectrum will be possible.

Our contemporary observations of the Crab Nebula show an excess rate 2.6 times larger than that for Mkn 421 and a spectral index of 2.7±0.3, from which a flux above 1 TeV of approx. $1.5 \times 10^{-11}$ cm$^{-2}$s$^{-1}$ is calculated. Thus the spectrum of Mkn 421 above 1 TeV appears to be steeper than the spectrum of the Crab Nebula and, within the statistical errors, the integral flux above 1 TeV from Mkn 421 is smaller than that from the Crab Nebula by a factor of 2.0±0.8.

## Acknowledgements

The HEGRA Collaboration thanks the Instituto de Astrofisica de Canarias for use of the HEGRA site at the Roque de los Muchachos and its facilities. This work was supported by the BMFT, the DFG and the CICYT.

## References

Aharonian F., et al., 1995, J. Phys. G21, 419
Biller S.D., et al., 1995, ApJ 445, 227
Fonseca V., 1995, in Proc. 24th ICRC, Rome, 1, 474
Kerrick A.D., et al., 1995, ApJ 438, L59
Konopelko A., et al., 1996, Astropart. Phys. 4, 199
Lamb R.C., et al., 1995, in Proc. 24th ICRC, Rome, 2, 491
Li T.P. & Ma Y.Q., 1983, ApJ 272, 317
Lin Y. C., et al., 1992, ApJ 401, L61
Macomb D.J., et al., 1995, ApJ 449, L99
Mirzoyan R., et al., 1994, Nucl. Instr. and Meth. A351, 513
Mirzoyan R., et al., 1995, in C. Cresti (ed.) Towards a Major Atmospheric Cherenkov Detector IV, Padua, 230
Mohanty G., et al., 1993, in Proc. 23rd ICRC, Calgary, 1, 440
Petry D., et al., 1995, in C. Cresti (ed.) Towards a Major Atmospheric Cherenkov Detector IV, Padua, 141
Petry D., 1996, PhD thesis, Munich, in preparation
Punch M., et al., 1992, Nature 358, 477
Rauterberg G. et al., 1995, in Proc. 24th ICRC, Rome, 3, 460
Reynolds P. T., et al., 1993, ApJ 404, 206
Wiedner C.A., 1994, in T. Kifune (ed.) Towards a Major Atmospheric Cherenkov Detector III, Tokyo, 119